\documentclass{aa}
\usepackage{graphicx}
\usepackage{natbib}
\usepackage{bm}
\graphicspath{{./fig/}{./png/}}
\newcommand{\EQ}{\begin{equation}}
\newcommand{\EN}{\end{equation}}
\newcommand{\EQA}{\begin{eqnarray}}
\newcommand{\ENA}{\end{eqnarray}}

\newcommand{\EEq}[1]{Equation~(\ref{#1})} %AB: for the beginning of a sentence
\newcommand{\Eq}[1]{Eq.~(\ref{#1})}

\newcommand{\Eqss}[2]{Eqs.~(\ref{#1})--(\ref{#2})}

\newcommand{\Sec}[1]{Sect.~\ref{#1}}

\newcommand{\FFig}[1]{Figure~\ref{#1}} %AB: for the beginning of a sentence
\newcommand{\Fig}[1]{Fig.~\ref{#1}}
\newcommand{\Figs}[2]{Figs.~\ref{#1} and \ref{#2}}

\newcommand{\bra}[1]{\langle #1\rangle}

\newcommand{\meanrho}{\overline{\rho}}

{}
{}
\newcommand{\meanFFFF}{\overline{\mbox{\boldmath ${\cal F}$}}{}}{}

\newcommand{\meanSSSS}{\overline{\mbox{\boldmath ${\mathsf S}$}} {}}

\newcommand{\meanSSS}{\overline{\mathsf{S}}}
{}
{}
{}
{}
{}
{}
{}
{}
\newcommand{\meanAA}{\overline{\mbox{\boldmath $A$}}{}}{}
\newcommand{\meanBB}{\overline{\mbox{\boldmath $B$}}{}}{}
{}
{}
{}
{}
{}
{}
{}
{}
\newcommand{\meanJJ}{\overline{\mbox{\boldmath $J$}}{}}{}
{}
\newcommand{\meanUU}{\overline{\bm{U}}}

\newcommand{\meanWW}{\overline{\mbox{\boldmath $W$}}{}}{}
{}
{}

\newcommand{\meanB}{\overline{B}}

\newcommand{\meanU}{\overline{U}}

\newcommand{\meanp}{\overline{p}}

{}

{}
{}

\newcommand{\zzz}{\hat{\mbox{\boldmath $z$}} {}}

\newcommand{\kk}{\bm{k}}
\newcommand{\pp}{\bm{p}}
\newcommand{\qq}{\bm{q}}
\newcommand{\xx}{\bm{x}}

\newcommand{\BB}{\mbox{\boldmath $B$} {}}

\newcommand{\grav}{\mbox{\boldmath $g$} {}}
\newcommand{\nab}{\mbox{\boldmath $\nabla$} {}}
\newcommand{\OO}{\bm{\Omega}}

\newcommand{\DD}{{\rm D} {}}

\newcommand{\dd}{{\rm d} {}}

\newcommand{\const}{{\rm const}  {}}

\def\Beqz{B_{\rm eq0}}

\def\Co{\mbox{\rm Co}}

\def\Rm{R_{\rm m}}

\def\Co{\mbox{\rm Co}}

\def\cs{c_{\rm s}}
\def\qp{q_{\rm p}}
\def\betastar{\beta_{\star}}
\def\betap{\beta_{\rm p}}

\def\vA{v_{\rm A}}

\def\kf{k_{\rm f}}

\def\urms{u_{\rm rms}}

\def\qpz{q_{\rm p0}}
\def\qp{q_{\rm p}}
\def\betap{\beta_{\rm p}}

\def\nut{\nu_{\rm t}}
\def\etat{\eta_{\rm t}}

\def\Beq{B_{\rm eq}}
\def\Peff{{\cal P}_{\rm eff}}

\def\half{{\textstyle{1\over2}}}

\def\onethird{{\textstyle{1\over3}}}

\newcommand{\etal}{et al.}
\newcommand{\yapj}[3]{ #1, {ApJ,} {#2}, #3}

\newcommand{\yapjl}[3]{ #1, {ApJ,} {#2}, #3}

\newcommand{\yan}[3]{ #1, {Astron.\ Nachr.,} {#2}, #3}

\newcommand{\yana}[3]{ #1, {A\&A,} {#2}, #3}

\newcommand{\ypfb}[3]{ #1, {Phys.\ Fluids B,} {#2}, #3}

\newcommand{\ysovl}[3]{ #1, {Sov.\ Astron.\ Lett.,} {#2}, #3}
\newcommand{\yjetp}[3]{ #1, {Sov.\ Phys.\ JETP,} {#2}, #3}

\newcommand{\ymn}[3]{ #1, {MNRAS,} {#2}, #3}
\newcommand{\ynat}[3]{ #1, {Nature,} {#2}, #3}

\newcommand{\ysph}[3]{ #1, {Solar Phys.,} {#2}, #3}

\newcommand{\ypre}[3]{ #1, {Phys.\ Rev.\ E,} {#2}, #3}

\newcommand{\yjour}[4]{ #1, {#2}, {#3}, #4}

\newcommand{\ybook}[3]{ #1, {#2} (#3)}

\newcommand{\sjour}[2]{ #1, {#2}, submitted}

\titlerunning{Negative magnetic pressure instability}
\authorrunning{I. R. Losada \etal}

\title{Rotational effects on the negative magnetic pressure instability}
\author{Illa R. Losada\inst{1,2}, A. Brandenburg\inst{3,4},
N. Kleeorin\inst{5,3}, Dhrubaditya Mitra\inst{3}, I. Rogachevskii\inst{5,3}
}
\institute{
Department of Astrophysics, Universidad de La Laguna,
38206 La Laguna (Tenerife), Spain
\and
Instituto de Astrof\'isica de Canarias, C/ V\'ia L\'actea, s/n, La Laguna,
Tenerife, Spain
\and
Nordita, Royal Institute of Technology and Stockholm University,
Roslagstullsbacken 23, 10691 Stockholm, Sweden
\and
Department of Astronomy, AlbaNova University Center,
Stockholm University, 10691 Stockholm, Sweden
\and
Department of Mechanical Engineering, Ben-Gurion University of the Negev,
POB 653, Beer-Sheva 84105, Israel
}
\date{\today,~ $ $Revision: 1.143 $ $}
\begin{document}

\abstract{
The surface layers of the Sun are strongly stratified.
In the presence of turbulence with a weak mean magnetic field,
a large-scale instability resulting in the formation of
nonuniform magnetic structures, can be excited
on the scale of many (more than ten) turbulent eddies (or convection cells).
This instability is caused by a negative
contribution of turbulence to the effective (mean-field)
magnetic pressure and has previously been discussed in connection
with the formation of active regions.
}{
We want to understand the effects of rotation on this instability
in both two and three dimensions.
}{
We use mean-field magnetohydrodynamics in a parameter regime in which
the properties of the negative effective magnetic pressure instability
have previously been found to agree with
properties of direct numerical simulations.
}{
We find that the instability is already suppressed for relatively
slow rotation with Coriolis numbers (i.e.\ inverse Rossby numbers) around 0.2.
The suppression is strongest at the equator.
In the nonlinear regime, we find traveling wave solutions with propagation
in the prograde direction at the equator with additional poleward
migration away from the equator.
}{
We speculate that the prograde rotation of the magnetic pattern near the
equator might be a possible explanation for the faster rotation speed
of magnetic tracers relative to the plasma velocity on the Sun.
In the bulk of the domain, kinetic and current helicities are negative
in the northern hemisphere and positive in the southern.
}

\maketitle

\section{Introduction}

In the outer parts of the Sun, energy is transported through turbulent
convection.
The thermodynamic aspects of this process are well understood through
mixing length theory \citep{Vit53}.
Also reasonably well understood is the partial conversion of kinetic
energy into magnetic energy via dynamo action \citep{Par79,ZRS83}.
Most remarkable is the possibility of generating magnetic fields on
much larger spatial and temporal scales
than the characteristic turbulence scales.
This has now been seen in many three-dimensional turbulence simulations
\citep{B01,BS05},
but the physics of this is best understood in terms of mean-field theory,
which encapsulates the effects of complex motions in terms of
effective equations for mean flow and mean magnetic field \citep{Mof78,Par79,KR80}.

The effects of stratification are usually only included to leading
order and often only in connection with rotation, because the two
together give rise to the famous $\alpha$ effect, which is able to
explain the generation of large-scale magnetic fields \citep{KR80}.
In recent years, however, a completely different effect arising from
strong stratification alone has received attention: the suppression
of turbulent pressure by a weak mean magnetic field.
This effect mimics a
negative effective (mean-field) magnetic pressure
owing to a negative contribution of turbulence to the mean
magnetic pressure.
Under suitable conditions, this leads
to the negative effective magnetic pressure instability
(NEMPI),
which can cause the formation of magnetic flux concentrations.
In turbulence simulations, this instability has
only been seen recently
\citep{BKKMR11}, because significant scale separation is needed
to overcome the effects of turbulent diffusion \citep{BKKR12}.
Mean-field considerations, however, have predicted the existence
of NEMPI for a long time \citep{KRR89,KRR90,KR94,KMR96,RK07,BKR10}.

One of the remarkable insights is that NEMPI can occur at any depth,
depending just on the value of the mean magnetic field strength.
However, for a domain of given depth the instability can only
occur in the location where the dependence of effective turbulent pressure
%on the ratio field strength to equipartition value has a negative slope.
%AB: the word "of" was missing
on the ratio of field strength to equipartition value has a negative slope.
Once this is obeyed, the only other necessary condition for NEMPI to occur
is that the turbulent diffusivity is
low enough.
In practice this means that there are enough turbulent eddies
within the domain of investigation \citep{BKKR12,KBKMR12b}.

Despite the potential importance of NEMPI, many additional effects have
not yet been explored.
The idea is that NEMPI would interact with the global dynamo producing
the large-scale magnetic field for NEMPI to act upon.
Thus, the field needs to be self-consistently generated.
Ideally, global geometry is needed, and such calculations should be
three-dimensional (3-D), because one expects flux concentrations not to
be two-dimensional (2-D) or axisymmetric.
New mean-field coefficients will appear in such a more general
case, and not much is known about them.
Nevertheless, although other terms may appear, it will be interesting
to investigate the evolution of NEMPI in more realistic cases with just
the leading term responsible for the instability.

The goal of the present paper is to include the effects of rotation
in NEMPI in a local Cartesian domain at a given latitude in the Sun.
To this end we
determine the dependence of growth rate and saturation
level of NEMPI on rotation rate and latitude, and to characterize
rotational effects on the resulting flux concentrations.
We restrict ourselves
to a mean-field treatment and denote averaged quantities by an overbar.
Furthermore, we make the assumption of an isothermal equation of state.
This is of course quite unrealistic, as far as applications to the Sun
are concerned.
However, it has been found earlier that NEMPI has similar properties
both for an isothermal layer with an isothermal equation of state
and a nearly isentropic one with the
more general perfect gas law \citep{Kapy12}.
Given that our knowledge of NEMPI is still rather limited, it is useful
to consider the new effects of rotation within the framework of the
conceptually simpler case of an isothermal layer.

We begin with the model equations,
discuss the linear theory of NEMPI in the presence of rotation,
and consider 2-D and 3-D numerical models.

\section{The model}

We consider here an isothermal equation of state with constant
sound speed $\cs$, so the mean gas pressure is $\meanp=\meanrho\cs^2$.
The evolution equations for mean velocity $\meanUU$, mean density $\meanrho$,
and mean vector potential $\meanAA$, are
\EQA
\label{dUmean}
{\DD\meanUU\over\DD t}&=&-2\OO\times\meanUU
-\cs^2\nab\ln\meanrho+\grav+\meanFFFF_{\rm M}+\meanFFFF_{\rm K},\\
{\DD\meanrho\over\DD t}&=&-\meanrho\nab\cdot\meanUU,\\
{\partial\meanAA\over\partial t}&=&\meanUU\times\meanBB-(\etat+\eta)\meanJJ,
\ENA
where $\DD/\DD t=\partial/\partial t+\meanUU\cdot\nab$
is the advective derivative,
$\etat$ and $\eta$ are turbulent and microscopic magnetic diffusivities,
$\grav=(0,0,-g)$ is the acceleration due to the gravity field,
\EQ
\meanFFFF_{\rm K}=(\nut+\nu)\left(\nabla^2\meanUU+\onethird\nab\nab\cdot\meanUU
+2\meanSSSS\nab\ln\meanrho\right)
\EN
is the total (turbulent plus microscopic) viscous force
with $\nut$ being the turbulent viscosity,
and $\meanSSS_{ij}=\half(\meanU_{i,j}+\meanU_{j,i})
-\onethird\delta_{ij}\nab\cdot\meanUU$
is the traceless rate of strain tensor of the mean flow.
The mean Lorentz force, $\meanFFFF_{\rm M}$, is given by
\EQ
\meanrho \, \meanFFFF_{\rm M} = \meanJJ\times\meanBB
+\half\nab(q_{\rm p}\meanBB^2),
\label{efforce}
\EN
where $\meanJJ=\nab\times\meanBB/\mu_0$
the mean current density,
$\mu_0$ is the vacuum permeability, and
the last term, $\half\nab(q_{\rm p}\meanBB^2)$, on the
righthand side of
\Eq{efforce} determines the turbulent contribution to
the mean Lorentz force. Following \cite{BKKR12} and \cite{KBKR12},
the function $\qp(\beta)$ is approximated by:
\EQ
\qp(\beta)={\betastar^2\over\betap^2+\beta^2},
\label{qp-apr}
\EN
where $\betastar$ and $\betap$ are constants,
$\beta=\meanB/\Beq$ is the modulus of the normalized mean
magnetic field, and $\Beq=\sqrt{\mu_0\rho}\, \urms$
the equipartition field strength.
The angular velocity vector $\OO$ is quantified by its scalar amplitude
$\Omega$ and colatitude $\theta$, such that
\EQ
\OO=\Omega\left(-\sin\theta, 0, \cos\theta\right).
\EN
In this arrangement, $z$ corresponds to radius,
$x$ to colatitude, and $y$ to azimuth.

Following the simplifying assumption of recent direct numerical
simulations of NEMPI \citep{BKKMR11}, we assume that the root-mean-square
turbulent velocity, $\urms$, is constant in space and time.
For an isothermal density stratification,
\EQ
\meanrho=\rho_0\exp(-z/H_\rho),
\label{rho}
\EN
where $H_\rho=\cs^2/g$ is the density scale height, we then have $\Beq(z)$.
To quantify the strength of the imposed field, we also define
$\Beqz=\Beq(z=0)$.
The value of $\urms$ is also related to the values of $\etat$ and $\nut$,
which we assume to be equal, with $\etat=\nut=\urms/3\kf$, where $\kf$
is the wavenumber of the energy-carrying eddies of the underlying turbulence.
This formula assumes that the relevant correlation time is $(\urms\kf)^{-1}$,
which has been shown to be fairly accurate \citep{SBS08}.

\section{Linear theory of NEMPI with rotation}
\label{LinTheo}

In this section we study the effect of rotation on the growth rate
of NEMPI.
Following earlier work \citep[e.g., the appendix of][]{KBKMR12b},
and for simplicity, we neglect dissipation processes,
use the anelastic approximation, $\nab\cdot\meanrho\meanUU=0$,
and assume that the density scale height $H_\rho=\const$.
We consider the equation of motion, ignoring
the $\meanUU\cdot\nab\meanUU$ nonlinearity,
\EQ
{\partial\meanUU(t,x,z)\over\partial t}=
-2\OO\times\meanUU -{1\over \meanrho} \nab p_{\rm tot}+\grav,
\label{A1}
\EN
where $p_{\rm tot}=\meanp + p_{\rm eff}$ is the total
pressure consisting of the sum of the mean gas pressure $\meanp$,
and the effective magnetic pressure,
$p_{\rm eff}=(1-\qp)\meanB^2\!/2$, where $\meanB=|\meanBB|$.
Here and elsewhere the vacuum permeability is set to unity.
We assume for simplicity that $\partial_y=0$, and that
the mean magnetic field
only has a $y$-component,
$\meanBB=(0,\meanB_y(x,z),0)$,
so the mean magnetic tension, $\meanBB\cdot\nab\meanBB$
in \Eq{A1} vanishes.

Taking twice the curl of \Eq{A1}, and noting further that
$\zzz\cdot\nab\times\nab\times\meanUU=-\Delta\meanU_z
+\nabla_z\nab\cdot\meanUU$,
we obtain
\begin{eqnarray}
&&{\partial\over\partial t} \left[\Delta \meanU_z
+ \nabla_z (\meanUU\cdot\nab \ln \meanrho)\right]=
- 2\OO\cdot\nab (\nab \times \meanUU)_z
\nonumber\\ && \quad\quad\quad
+\nabla_x\biggl[\left(\nabla_z{p_{\rm tot}\over\meanrho}\right)
{\nabla_x\meanrho\over\meanrho}
-\left(\nabla_x{p_{\rm tot}\over\meanrho}\right){\nabla_z\meanrho\over\meanrho}
\biggr],
\label{A2}
\end{eqnarray}
where we have used the anelastic approximation in the form
$\nab\cdot\meanUU=-\meanUU\cdot\nab\ln\meanrho$ and the fact
that under the curl the gradient can be moved to $\meanrho$.
We have also taken into account that $\Omega_{y}=0$
and have used Eq.~(30) of \cite{KBKMR12b} to relate the
double curl
of $(\nab p_{\rm tot})/\meanrho$ to the last term in \Eq{A2}.
The first term on the
righthand
side of \Eq{A2} for $\meanU_z$
is proportional to $(\nab \times \meanUU)_z$.
Taking the $z$ component of the curl of \Eq{A1} we obtain
the following equation for $(\nab \times \meanUU)_z$:
\begin{eqnarray}
{\partial\over\partial t} (\nab \times \meanUU)_z=
2 \left(\OO\cdot\nab  - {\Omega_{z}\over H_\rho} \right) \meanU_z.
\label{B1}
\end{eqnarray}
The induction equation for $\meanB_y(x,z)$ is given by
\EQ
{\DD\meanB_y\over\DD t}=-\meanB_y\nab\cdot\meanUU,
\label{B2}
\EN
where $\DD/\DD t=\partial/\partial+\meanUU\cdot\nab$
is the advective derivative.
For a magnetic field with only a $y$-component, but
$\partial/\partial y=0$, there is no stretching term, so
there is no term of the form $\meanBB\cdot\nab\meanUU$.

We linearize \Eqss{A2}{B2}, indicating small changes by $\delta$.
We consider an equilibrium with a constant magnetic field of the
form $(0,B_0,0)$, a zero mean velocity, and
the fluid density as given by \Eq{rho}.
We take into account that
the function $\qp=\qp(\beta)$ depends both on $\meanB$
and on $\meanrho$, which implies that \citep{KBKMR12b}
\EQ
\delta\left({p_{\rm tot}\over\meanrho}\right)=\half\vA^2
\left(1-\qp-{\dd\qp\over \dd\ln\beta^2}\right)
\left(2{\delta\meanB_y\over B_0}-{\delta\meanrho\over\meanrho}\right),
\label{A2b}
\EN
while
\EQ
\nabla_z\left({p_{\rm tot}\over\meanrho}\right)=\half\vA^2
\left(1-\qp-{\dd\qp\over \dd\ln\beta^2}\right)
{1\over H_\rho}.
\label{A2c}
\EN
The linearized system of equations reads as
\begin{eqnarray}
{\partial\over\partial t} \left(\Delta - {1\over H_\rho}\nabla_z\right)
\delta\meanU_z&=&2{\vA^2\over H_\rho}{\dd\Peff\over\dd\beta^2}
{\nabla_x^2\delta\meanB_y\over B_0}
\nonumber\\
&&- 2\OO\cdot\nab (\nab \times \delta\meanUU)_z,
\label{B3}\\
{\partial\over\partial t} (\nab \times \delta\meanUU)_z&=&
2 \left(\OO\cdot\nab  - {\Omega_{z}\over H_\rho} \right) \delta\meanU_z,
\label{B4}\\
{\partial\delta\meanB_{y}\over\partial t}&=&-B_0{\delta\meanU_{z}\over H_\rho},
\label{B5}
\end{eqnarray}
where $\Peff(\beta)=\half\left[1-\qp(\beta)\right]\beta^2$ is the
effective magnetic pressure normalized by the local value of $\Beq^2$.

Introducing a new variable $V_z=\sqrt{\meanrho}\,\delta\meanU_z$
in \Eqss{B3}{B5} and after simple transformations
we arrive at the following equation for one variable $V_z$:
\begin{eqnarray}
{\partial^2\over\partial t^2} \left(\Delta - {1 \over 4 H_\rho^2}\right) V_z
+ \left((2\OO\cdot\nab)^2 - {\Omega_{z}^2\over H_\rho^2} \right) V_z
&\!=\!&\lambda^2_0 \nabla_x^2 V_z,
\nonumber\\ &&
\label{A6}
\end{eqnarray}
where
\EQ
\lambda_0^2(z)= -2 {\vA^2(z) \over H_\rho^2} \,
{\dd\Peff(z)\over\dd\beta^2}.
\label{AA11}
\EN

In the WKB approximation, which is valid when $k_z \, H_\rho \gg 1$,
i.e., when the characteristic scale of the spatial variation of the
perturbations of the magnetic and velocity fields are much smaller than the
density height length, $H_\rho$, the growth rate of the large-scale instability
(NEMPI) is given by
\EQ
\lambda= \left[\lambda_0^2 \,{k_x^2 \over k^2} - \omega_{\rm inert}^2\right]^{1/2},
\label{B6}
\EN
where $\omega_{\rm inert} = 2\OO\cdot{\hat{\bm k}}$ is the frequency of
the inertial waves.
Here, ${\hat{\bm k}}={\bm k}/k$ is the unit vector of ${\bm k}$.
A necessary condition for the instability is
\EQ
{\dd\Peff\over\dd\beta^2} < 0 .
\label{A10}
\EN
NEMPI can be excited even in a uniform mean magnetic field,
and the source of free energy of the instability is provided by the
small-scale turbulence.
In contrast, the free energy in Parker's magnetic buoyancy
instability \citep{Par66} or in the interchange instability
\citep{Tse60,PR82} is drawn from the gravitational field.
Both instabilities are excited in a plasma when the characteristic scale
of variations in the original horizontal magnetic field
is smaller than the density scale height.
As seen from Eq.~(\ref{B6}),
$\lambda$ is either real or purely imaginary, so no complex
eigenvalues are possible, as would be required for growing oscillatory
solutions.

Without rotation the growth rate of the large-scale instability is
\citep{KMR93,RK07,KBKMR12c}
\EQ
\lambda= \lambda_0 \, {k_x \over k}.
\label{A11}
\EN
The rotation reduces the growth rate of NEMPI,
which can be excited when $k_x/k>\omega_{\rm inert}/\lambda_0$
and $\dd\Peff / \dd\beta^2 < 0$.
In the opposite case, $k_x/k<\omega_{\rm inert}/\lambda_0$,
the large-scale instability is not excited,
while the frequency of the inertial waves is reduced
by the effective negative magnetic pressure.

\begin{figure}[t!]\begin{center}
\includegraphics[width=\columnwidth]{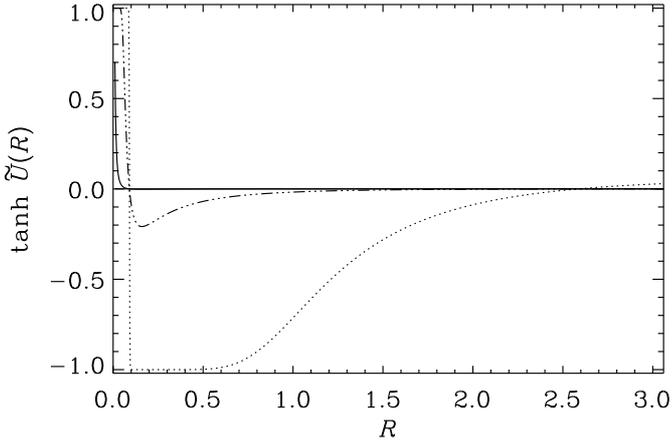}
\end{center}\caption[]{
$\tanh\tilde{U}(R)$ for
$\tilde\lambda\equiv\lambda/\lambda_\ast=0.02$,
$\theta=0$, and $\Omega=0.01$ (dotted line),
0.1 (dashed-dotted line), and 1 (solid line).
}\label{potential}\end{figure}

For an arbitrary vertical inhomogeneity of the density,
we seek a solution
to \Eq{A6} in the form $V_z(t,x,z) = V(z)
\exp(\lambda t+ i k_x \,x)$ and obtain an eigenvalue problem
\begin{eqnarray}
\biggl[\nabla_z^2 + {8 \Omega_{x} \Omega_{z} \over
\lambda^2 + 4 \Omega_{z}^2} \, i k_x \, \nabla_z
- \Lambda^2 \, k_x^{2}
-{1\over4H_\rho^2}\biggr] V(z) = 0,
\label{B8}
\end{eqnarray}
where
\begin{eqnarray}
\Lambda^2={\lambda^2 - \lambda_0^{2}(z) + 4 \Omega_{x}^2
\over \lambda^2 + 4 \Omega_{z}^2},
\end{eqnarray}
and $\lambda$ is the eigenvalue.
\EEq{B8} can be reduced to the Schr\"odinger type equation,
$\Psi''-\tilde{U}(R) \, \Psi=0$, via the transformation
\begin{eqnarray}
&&\Psi(R)=\sqrt{R} \, V(z) \, \exp \left(i {4 \Omega_{x} \Omega_{z} \over
\lambda^2 + 4 \Omega_{z}^2} k_x  z \right) ,
\label{B9}\\
&& R(z) ={v_{A0}^2 \over u_{\rm rms}^2 \betap^2} \,e^{z/H_\rho} ,
\label{A7}
\end{eqnarray}
where
$v_{\rm A0}=B_0/\sqrt{\meanrho_0}$ is the Alfv\'en speed based on the
averaged density, the potential $\tilde U(R)$ is
\begin{eqnarray}
\tilde U(R)&=& {k_x^2 H_\rho^2 \over R \, (\lambda^2 + 4 \Omega_{z}^2)}
\biggl[{\lambda^2 \over R} \left({\lambda^2 + 4 \Omega^2
\over \lambda^2 + 4 \Omega_{z}^2} \right)
\nonumber\\
&&+ {u_{\rm rms}^2 \, \betap^2 \over H_\rho^2}
\left(1- {\qpz\over (1+R)^2}\right)\biggr] ,
\label{A8}
\end{eqnarray}
and we have used \Eq{qp-apr} for $\qp$ with $\betastar=\betap \sqrt{\qpz}$
and $\qpz=\qp(\beta=0)$.
As follows from \Eq{A8},
the potential, $\tilde U(R)$, is positive for $R\to 0$ and $R\to \infty$.
Therefore, for the existence of the instability, the potential should
have a negative minimum. This is possible
when $\qpz > (1+R)^2$.
When the potential $\tilde U(R)$ has a negative minimum, there are two points
$R_1$ and $R_2$ (the so-called turning points) in which $\tilde U(R=R_{1,2})=0$.
\FFig{potential} shows $\tanh\tilde{U}(R)$ for different values of
$\Omega$.
This representation allows us to distinguish
the behavior for low values of $\tilde{U}(R)$.

\begin{figure}[t!]\begin{center}
\includegraphics[width=\columnwidth]{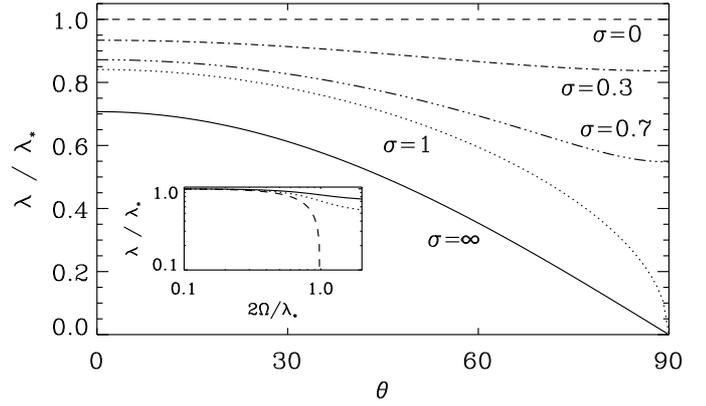}
\end{center}\caption[]{
Theoretical dependence of $\lambda/\lambda_\ast$ on $\theta$
for different values of $\sigma$ using \Eq{AA13}.
The inset shows the dependence of $\lambda/\lambda_\ast$ on
$2\Omega/\lambda_\ast=\sigma^{1/2}$ for $\theta=0^\circ$
(solid), $45^\circ$ (dotted), and $90^\circ$ (dashed).
}\label{plam_sigma}\end{figure}

Using \Eq{A8} and the condition $\tilde U(R=R_{1,2})=0$,
we estimate the maximum growth rate of the instability as
\EQ
\lambda={1\over\sqrt{2}}\left[\lambda_\ast^2 - 4\Omega^2 + \left[(\lambda_\ast^2
- 4\Omega^2)^2 + 8\Omega_{z}^2 \lambda_\ast^2\right]^{1/2}\right]^{1/2},
\label{A14}
\EN
where
\EQ
\lambda_\ast= {\betastar \,u_{\rm rms} \over H_\rho} \,
{\left[R_1 R_2 (2+R_1+R_2) \right]^{1/2}\over (1+R_1)(1+R_2)} .
\label{A12}
\EN
By defining $\sigma=4\Omega^2/\lambda_\ast^2$, \Eq{A14}
can also be written as
\EQ
\lambda/\lambda_\ast={1\over\sqrt{2}}\left[1-\sigma
+\left(1-2\sigma \sin^2 \theta+\sigma^2\right)^{1/2}\right]^{1/2}.
\label{AA13}
\EN
For $\sigma \gg 1$, we obtain
$\lambda/\lambda_\ast=\cos \theta/\sqrt{2}$, which is
independent of the value of $\sigma$.
In \Fig{plam_sigma} we plot the dependence of $\lambda/\lambda_\ast$
on $\theta$ for different values of $\sigma$
and on $2\Omega/\lambda_\ast=\sigma^{1/2}$ for different values of
$\theta$ (inset).

Unfortunately, the asymptotic analysis does not allow full information
about the system.
Therefore we turn in the following to numerical simulations of the
full 2-D and 3-D mean-field equations.

\section{Numerical results}

In this section we discuss numerical mean-field
modeling.
We consider computational domains of size $L^2$ or $L^3$ with periodic
boundary conditions in the horizontal direction(s) and stress-free perfect
conductor boundary conditions in the vertical direction.
The smallest wavenumber that fits horizontally into the domain has the
wavenumber $k_1=2\pi/L$.
The numerical simulations are performed with the {\sc Pencil Code}%
\footnote{{\tt http://pencil-code.googlecode.com}},
which uses sixth-order explicit finite differences in space and a
third-order accurate time stepping method \citep{BD02}.
As units of length we use $k_1^{-1}$, and time is measured
in units of $(\cs k_1)^{-1}$.

An important nondimensional parameter is the Coriolis number,
$\Co=2\Omega/\urms\kf$.
Using $\kf=\urms/3\etat$, we can express this in terms of
the parameter $C_\Omega=\Omega/\etat k_1^2$, which is often used in
mean-field dynamo theory.
Thus, we have
\EQ
\Co=6\etat\Omega/\urms^2=6\,(\etat k_1/\urms)^2C_\Omega.
\EN
Motivated by the analytic results of the previous section we normalize
the growth rate of the instability
alternatively by a quantity $\lambda_{*0}\equiv\betastar\urms/H_\rho$.
In the following we take $\urms/\cs=0.1$.
Furthermore,
we use $\nut=\etat=10^{-3}\cs/\kf$, so that $\kf H_\rho\approx33$
and $\etat k_1/\urms=10^{-2}$.
This also means that for $\Omega=0.01$, for example, we have
$2\Omega/\lambda_{\ast0}=0.27$ and $\Co=0.006$.

For the models presented below, we use $\qpz=20$ and $\betap=0.167$,
which corresponds to $\betastar=0.75$, and is appropriate for the
parameter regime in which $\Rm\approx18$ and $\kf/k_1=30$ \citep{KBKMR12c}.
We use either $B_0/\Beqz=0.1$ or $0.05$.
We recall, however, that the growth rate does not depend on
this choice, provided the bulk of the eigenfunction fits into the domain,
which is the case here
for both values of $B_0$.
For the lower value of $B_0$ the maximum of the magnetic structures
(i.e., the maximum of the eigenfunction in $z$)
is slightly higher up in the domain, but in both cases the maximum is
contained within the domain.

\begin{figure}[t!]\begin{center}
\includegraphics[width=\columnwidth]{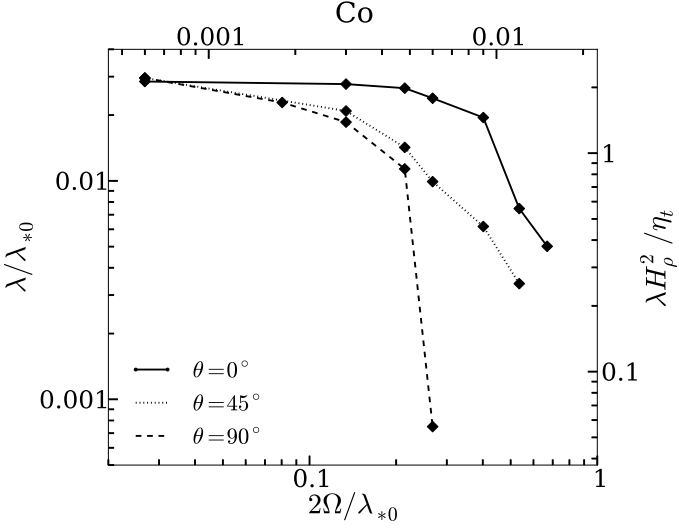}
\end{center}\caption[]{
Dependence of $\lambda/\lambda_{\ast0}$ on $2\Omega/\lambda_{\ast0}$ for
three values of $\theta$ for 2-D simulations with $B_0/\Beqz=0.1$.
}\label{fig:growth_om2D}\end{figure}

We discuss first the $\Omega$ and $\theta$ dependence of 2-D and 3-D solutions.
Using $\theta=0^\circ$, $45^\circ$, and $90^\circ$,
corresponding to $90^\circ$, $45^\circ$, and $0^\circ$ latitude,
we find that NEMPI is suppressed for rotation rates around
$\Omega\approx0.01\cs k_1$ and $0.025$ in 2-D and 3-D, as can be seen in
\Figs{fig:growth_om2D}{presults_Omdep}.
This corresponds to $\Co=0.006$ and $0.015$, which are remarkably
low values.
We note a similar behavior
in 2-D and 3-D: NEMPI is suppressed for even
lower values of $2\Omega/\lambda_{\ast0}$ as $\theta$ increases.
Moreover, there is qualitative agreement between the results of mean-field
simulations and the predictions based on asymptotic analysis, even though
in the former case we normalized by $\lambda_{\ast0}$, while in the latter
we normalized by $\lambda_{\ast}$; see \Eq{AA13}.

\begin{figure}[t!]\begin{center}
\includegraphics[width=\columnwidth]{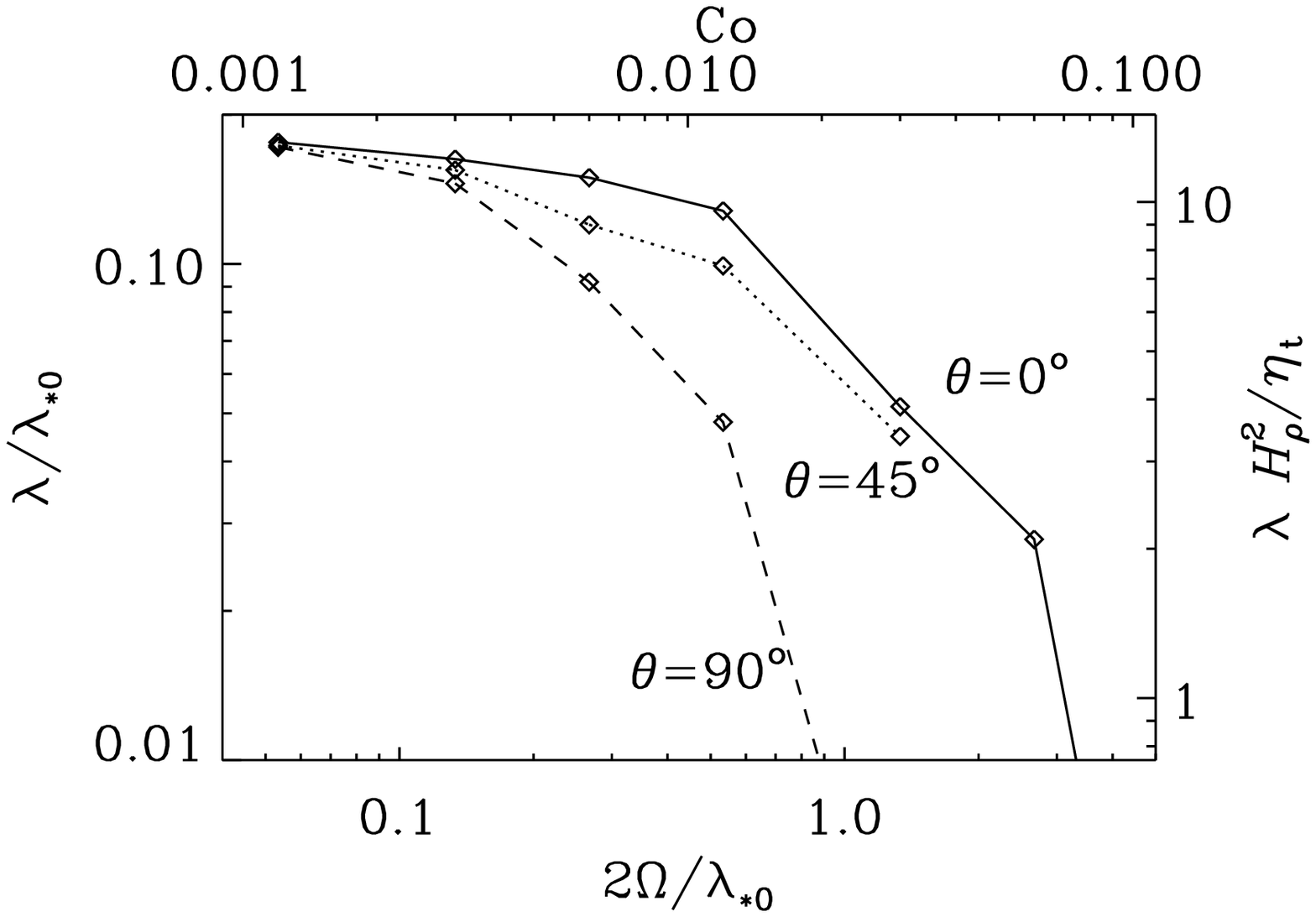}
\end{center}\caption[]{
Dependence of $\lambda/\lambda_{\ast0}$ on $2\Omega/\lambda_{\ast0}$ for three
values of $\theta$ for
3-D simulations with $B_0/\Beqz=0.05$.
}\label{presults_Omdep}\end{figure}

\begin{figure}[t!]\begin{center}
\includegraphics[width=\columnwidth]{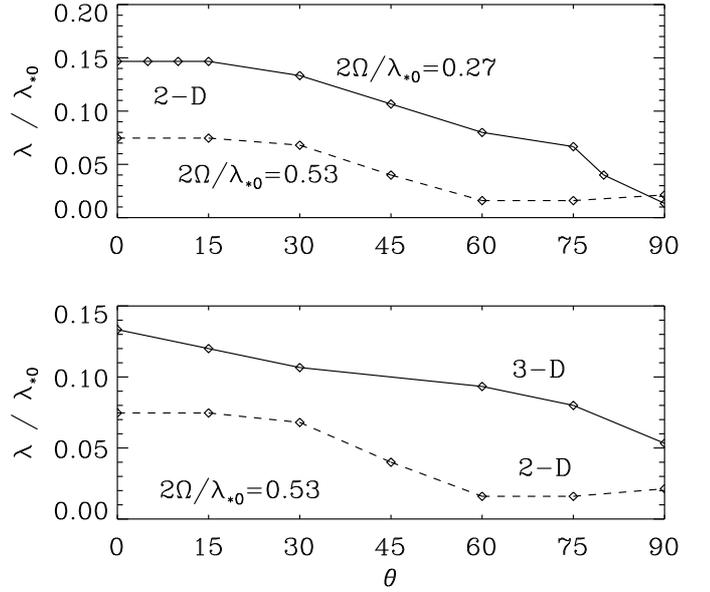}
\end{center}\caption[]{
Dependence of $\lambda/\lambda_{\ast0}$ on $\theta$ for two values of
$2\Omega_0/\lambda_{*0}$ in 2-D (upper panel)
and comparison of 2-D and 3-D cases (lower panel).
}\label{presults_Om02_theta_3d}\end{figure}

Next, we vary $\theta$.
As expected from the results of \Sec{LinTheo},
and as already seen in \Figs{fig:growth_om2D}{presults_Omdep},
the largest growth rates occur at the poles $(\theta=0^\circ)$,
and NEMPI is the most strongly suppressed at the equator.
The growth rate as a function of $\theta$ is given in
\Fig{presults_Om02_theta_3d} for two values of $2\Omega/\lambda_{\ast0}$,
showing a minimum at $\theta=90^\circ$ (i.e., at the equator).
In the upper panel of \Fig{presults_Om02_theta_3d}, we have used
2-D results, i.e.\ we
restricted ourselves to solutions
with $\partial/\partial y=0$, as was also done in \Sec{LinTheo}.
However, this is only an approximation
of the fully 3-D case.
The usefulness of this restriction can be assessed by comparing
2-D and 3-D results; see the lower panel of
\Fig{presults_Om02_theta_3d}.
While the $\theta$ dependence is roughly similar in the
2-D and 3-D cases, the growth rates are by at least a factor of
two lower in the 2-D case.

\begin{figure}[t!]\begin{center}
\includegraphics[width=\columnwidth]{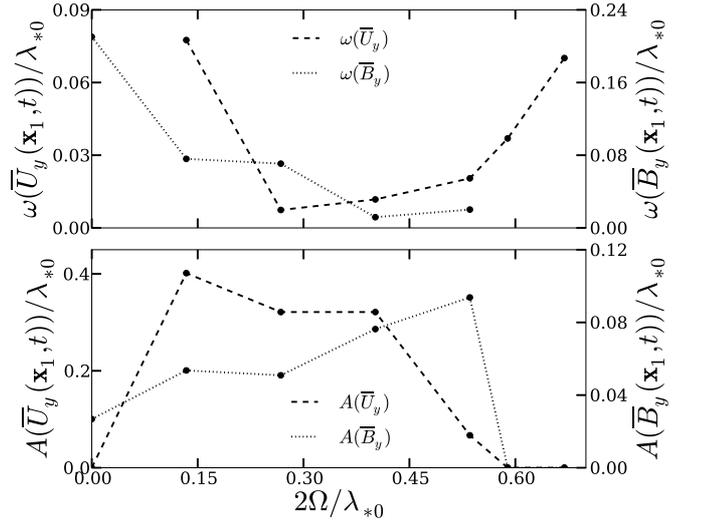}
\end{center}\caption[]{Frequency and amplitude as a function of $\Omega$ for
$\theta=60^\circ$ and $B_0/\Beqz=0.1$ in the saturated regime.
}\label{fig:frec_omega}\end{figure}

To determine the oscillatory frequency, we consider
the values of $\meanU_y(\xx_1,t)$ and $\meanB_y(\xx_1,t)$ at a fixed point
$\xx_1$ within the domain.
As can be seen in \Figs{fig:frec_omega}{fig:frec_theta},
their frequency and amplitude depend
on both $\Omega$ and $\theta$.
The oscillations are not always harmonic ones, and can be irregular with
variable periods, making the period determination more difficult.
Nevertheless, the frequencies for $\meanU_y$ and $\meanB_y$ are similar
over broad parameter ranges.
For $\Omega_0/\lambda_{*0}>0.25$ at $\theta=60^\circ$, NEMPI is no longer
excited, but there are still oscillations in
$\meanU_y(\xx_1,t)$, which must then have some other cause.
We find a substantial variation
in the amplitude for the maximum
growth rate for $\Omega=0.01$ and $\Omega=0.02$.
(The high frequency in $\meanU_y$ and $\meanB_y$ in
\Fig{fig:frec_omega} corresponds to a random small-amplitude change.)
The frequency of the oscillations is very
low at the poles, but it reaches a maximum
at $\theta=45$ and decreases again toward the equator.

\begin{figure}[t!]\begin{center}
\includegraphics[width=\columnwidth]{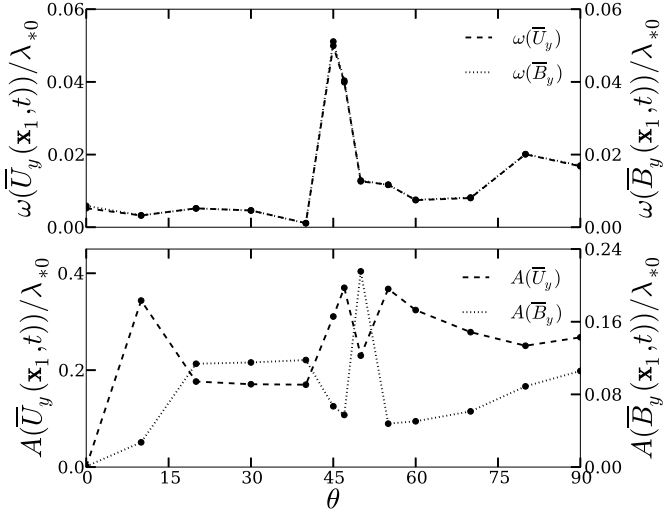}
\end{center}\caption[]{Frequency and amplitude $\theta$ dependence for
$\Omega=0.01$ and $B_0/\Beqz=0.1$.
}\label{fig:frec_theta}\end{figure}

In summary, the oscillation frequency decreases (and the period increases)
for faster rotation as the growth rate diminishes.
Furthermore, the oscillation frequency is systematically
lower at low
latitudes (below $45^\circ$) and higher closer to the poles.
We recall that these oscillations occur only in the nonlinear regime,
so no meaningful comparison with linear theory is possible.

\begin{figure}[t!]\begin{center}
\includegraphics[width=\columnwidth]{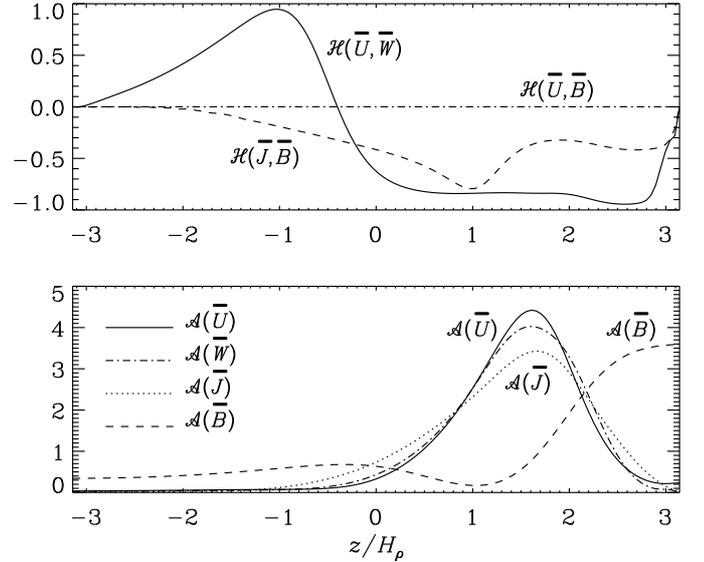}
\end{center}\caption[]{
Dependence of various relative helicities and relative amplitudes on $z$
for the case with $\theta=0^\circ$ and $\Co=0.03$.
}\label{phel}\end{figure}

Given the combined presence of rotation and stratification, we expect
the resulting velocity and magnetic fields to be helical.
We plot relative kinetic, current, and
cross helicities in the upper panel of \Fig{phel}.
These are here abbreviated in terms of the function
\EQ
{\cal H}(\pp,\qq)=\bra{\pp\cdot\qq}/\sqrt{\bra{\pp^2}\bra{\qq^2}},
\EN
where $\pp$ and $\qq$ are two arbitrary vectors.
Here, $\bra{\cdot}$ denotes $xy$ averaging.
The relative kinetic helicity, ${\cal H}(\meanWW,\meanUU)$,
where $\meanWW=\nab\times\meanUU$ is the mean vorticity, varies between
nearly $+1$ in the lower part and $-1$ in the upper part.
This change of sign is familiar from laminar convection where upwellings expand to
produce negative helicity in the upper parts, and downwellings also expand
as they hit the bottom of the domain \citep[e.g.][]{BNPST90}.
However, in the lower part of the domain
both $\meanUU$ and $\meanWW$ are relatively small,
as can be seen by considering their relative amplitudes,
${\cal A}(\meanUU)$ and ${\cal A}(\meanWW)$, where
\EQ
{\cal A}(\pp)=\bra{\pp^2}/\bra{\bra{\pp^2}},
\EN
with $\bra{\bra{\cdot}}$ being defined as volume averages.

It will be important to compare the present predictions of large-scale
kinetic and magnetic helicity production with results from future DNS.
One might expect differences between the two, because our current
mean-field models ignore turbulent transport coefficients that are
associated with helicity; see the discussion at the end of \cite{KBKR12b}.

\begin{figure}[t!]\begin{center}
\includegraphics[width=\columnwidth]{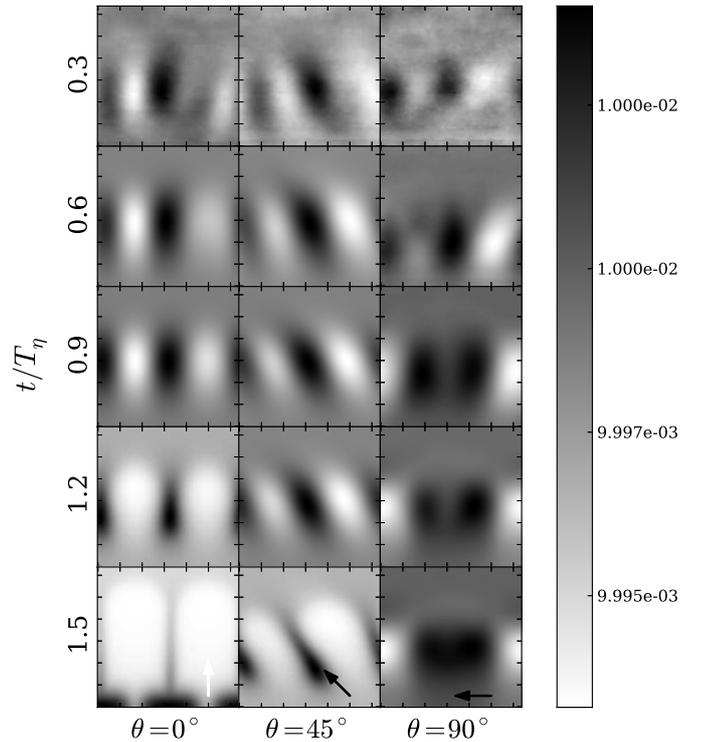}
\end{center}\caption[]{
Evolution of $\meanB_y$ in the $xz$ plane in a 2-D simulation
for $\Omega_0=0.01$ (corresponding to $\Co=0.006$)
and $B_0/\Beqz=0.1$ for
$\theta=0^\circ$, $\theta=45^\circ$, and $\theta=90^\circ$
near the time when the instability saturates.
The direction of $\OO$ is indicated in the last row.
}\label{fig:time_evolution}\end{figure}

\begin{figure*}[t!]\begin{center}
\includegraphics[width=\textwidth]{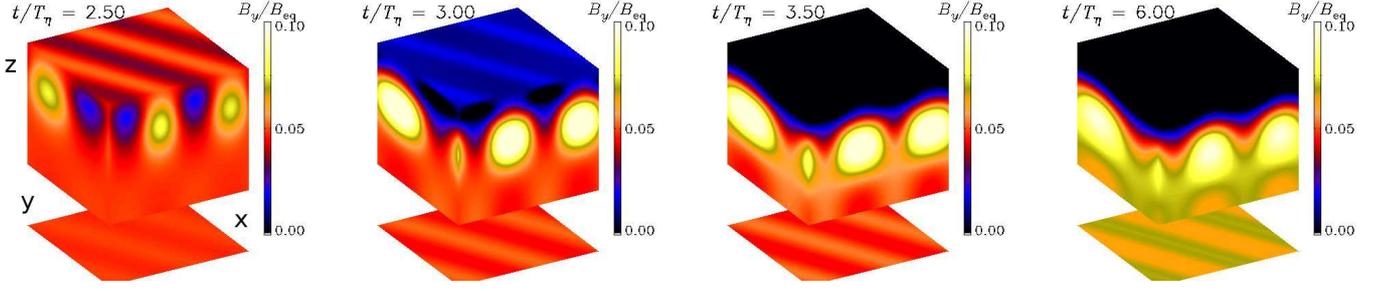}
\end{center}\caption[]{
Visualization of $B_y$ on the periphery of the computational domain
for 4 times (normalized in terms of $T_\eta$) during the nonlinear
stage of the instability for $\theta=0^\circ$ (corresponding to the north pole)
and $\Co=0.03$, corresponding to $2\Omega/\lambda_{\ast0}\approx1.3$.
Time is here given in units of $T_\eta=(\etat k_1^2)^{-1}$.
}\label{B_Om05_b005_th0}\end{figure*}

\begin{figure*}[t!]\begin{center}
\includegraphics[width=\textwidth]{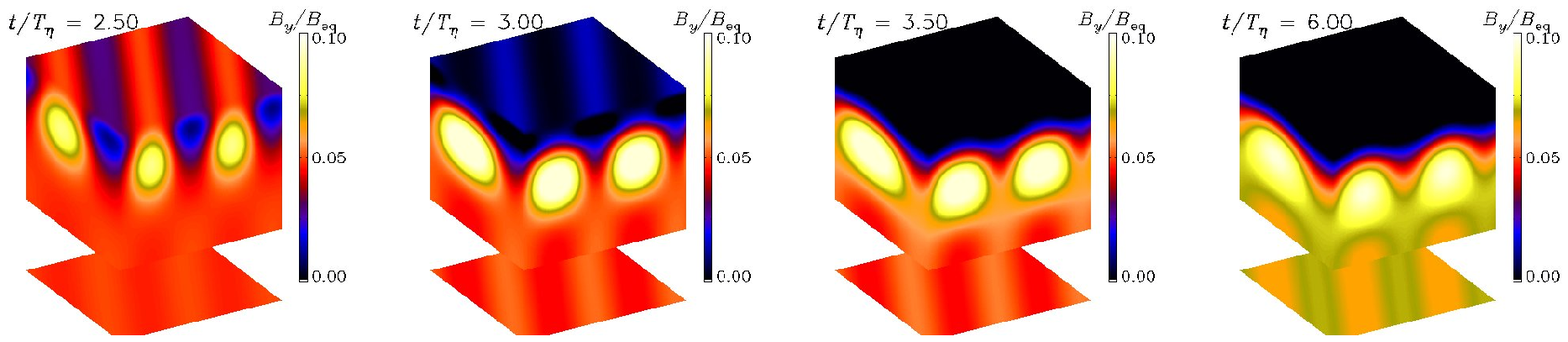}
\end{center}\caption[]{
Same as \Fig{B_Om05_b005_th0}, but for a negative value of $\Omega$,
i.e., $\Co=-0.03$, corresponding to $2\Omega/\lambda_{\ast0}\approx-1.3$.
}\label{B_Omm05_b005_th0}\end{figure*}

We finally turn to the spatial structure of NEMPI.
In \Fig{fig:time_evolution} we compare $\meanB_y$ at different times
and latitudes for the 2-D runs.
In the exponentially growing phase of NEMPI,
the structures do not propagate (or move only very slowly).
Traveling wave solutions occur mainly in
a later stage of NEMPI, i.e.,
in the saturated regime.
Next, we consider the 3-D case.
In \Fig{B_Om05_b005_th0} we show visualizations of the magnetic field on the
periphery of the computational domain for
four different times for $\theta=0$.
Magnetic structures are inclined in the $xy$ plane.
This is a direct result of rotation.
As expected, the inclination is opposite for negative values
of $\Omega$; see \Fig{B_Omm05_b005_th0}.
The modulus of the inclination angle is about $30^\circ$,
corresponding to 0.5 radians, which is not compatible with
the value of $\Co\approx0.03$, but it is closer to the value of
$\Omega/\lambda_{\ast0}\approx0.65$.
However, in this connection we should stress that we have imposed
periodic boundary conditions in the $y$ direction, which means
that the inclination angles
only change in discrete steps.
In the 2-D runs, shown in \Fig{fig:time_evolution}, no inclination
in the $xy$ plane is possible at all.


\begin{figure*}[t!]\begin{center}
\includegraphics[width=\textwidth]{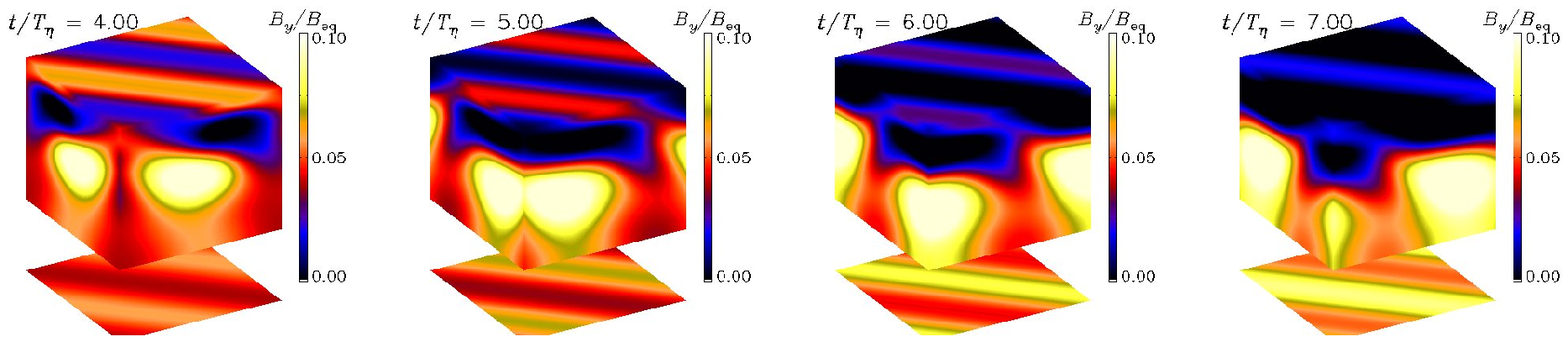}
\end{center}\caption[]{
Visualization of $B_y$ on the periphery of the computational domain
for 4 times (normalized in terms of $T_\eta$) during the nonlinear
stage of the instability for $\theta=45^\circ$ and $
\Co=0.03$, corresponding to $2\Omega/\lambda_{\ast0}\approx1.3$.
}\label{B}\end{figure*}

\begin{figure*}[t!]\begin{center}
\includegraphics[width=\textwidth]{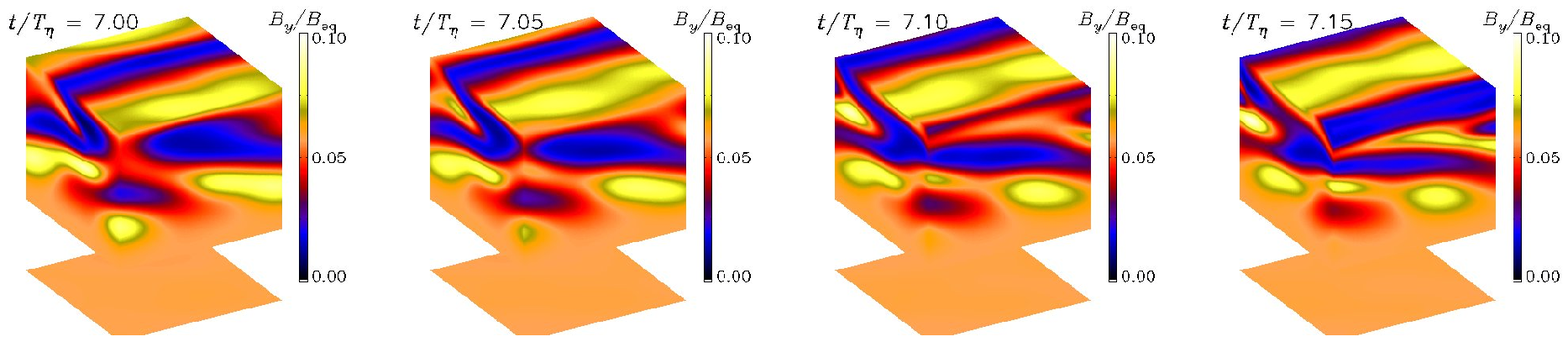}
\end{center}\caption[]{
Visualization of $B_y$ on the periphery of the computational domain
for 4 times (normalized in terms of $T_\eta$) during the nonlinear
stage of the instability for $\theta=90^\circ$ (corresponding to the equator)
and $\Co=0.013$, corresponding to $2\Omega/\lambda_{\ast0}\approx0.5$.
}\label{B_Om02_b005_th90}\end{figure*}

Returning to the case of positive values of $\Omega$, but $\theta\neq0$,
we note a slow migration of the magnetic pattern to the left (here for
$\theta=45^\circ$), corresponding to poleward migration; see \Fig{B}.
Also the field is still tilted in the $xy$ plane.
Finally, for $\theta=90^\circ$ we see that the pattern speed corresponds to
prograde motion; see \Fig{B_Om02_b005_th90}.

\section{Conclusions}

Although the physical reality of NEMPI
has recently been confirmed by direct numerical simulations,
its potential role
in producing large-scale magnetic structures in
the Sun is still unclear.
This paper begins the task of investigating its
properties under conditions that are astrophysically important.
Rotation is ubiquitous and clearly important in the Sun.
The present work has now shown that
the instability is suppressed already for rather slow rotation.
This is rather surprising, because rotational effects normally become
significant only when $\Omega$ is comparable
to the inverse turnover
time, which is defined here as $\urms\kf$.
The instability growth rate scale might explain this behaviour, since it is
closer to the turbulent diffusive time than to the inverse turnover, which
is faster by the square of the scale separation ratio \citep{BKKMR11}.
However,
our work now suggests that this is not quite right either
and that the correct answer might be something in between.
Indeed, we find here that growth rate and critical rotation rate are
close to the parameter $\lambda_{\ast0}=\betastar\urms/H_\rho$, which
can be smaller than the aforementioned turnover time by a factor of 40,
although in solar convection, where $\kf H_\rho\approx2.4$ \citep{KBKMR12c}
and $\betastar\approx0.23$ \citep{KBKMR12b}, it is estimated to be only
$\approx10$ times smaller.

The suppression is strongest at the equator, where
$\OO$ is perpendicular to the direction of the gravity field,
i.e., $\OO\cdot\grav=0$,
and less strong at the poles where $\OO$ and $\grav$ are either
parallel (south pole) or antiparallel (north pole).
In the absence of rotation, the mean magnetic field only varies
in a plane that is normal to the direction of the imposed mean
magnetic field, i.e., $\kk\cdot\BB_0=0$, where $\kk$ stands
for the wave vector of the resulting flow and magnetic field.
However, in the presence of rotation the orientation
of this plane changes such that now
$\kk\cdot(\BB_0+\lambda_{\ast0}^{-1}\OO\times\BB_0)=0$.

At intermediate latitudes, i.e., when the angle spanned by $\OO$ and $\grav$ is
in the range
of $0^\circ$ to $90^\circ$ colatitude, the magnetic field pattern
propagates slowly in the negative $x$ direction, corresponding
to poleward migration.
The significance of this result is unclear.
Had it been equatorward migration, one might have been tempted
to associate this with the equatorward migration of the
magnetic flux belts in the Sun from which sunspots emerge.
On the other hand, at the equator this migration corresponds to
prograde rotation, which is a clear effect seen in the Sun where
magnetic tracers are seen to rotate faster than the ambient plasma,
i.e., in the prograde direction \citep{GDS03}.
Even sunspots rotate faster than the gas itself \citep{PT98}.

One of our goals for future work is to verify the present findings
in direct numerical simulations.
Such simulations would also allow us to determine new turbulent transport
coefficients, similar to the $\qp$ parameter invoked in the present study.
Such additional parameters yield new effects, some of which could be
important for applications to the Sun.

Finally, we end with a comment on
the issue of scale separation.
As discussed above, in solar mixing length theory, the correlation length of the
turbulent eddies is expected to scale with the pressure
scale height such that $\kf H_\rho$ is constant and about 2.4 \citep{KBKMR12c}.
Theoretical considerations have shown further that the growth rate of NEMPI
is proportional to $\kf H_\rho$.
Since rotation is known to decrease the size of the turbulent eddies,
i.e., to increase the value of $\kf$, one might be tempted to speculate
that rotation could even enhance the growth rate of NEMPI.
However, in view of the present results, this
now seems unlikely.

\begin{acknowledgements}
We thank the anonymous referee for making many useful suggestions that have
improved the presentation of our results.
Illa R.\ Losada was supported by PhD Grant `Beca de Investigaci\'on
CajaCanarias para Postgradua­dos 2011'.
This work was supported in part by
the European Research Council under the
AstroDyn Research Project No.\ 227952,
by the National Science Foundation under Grant No.\ NSF PHY05-51164 (AB),
by EU COST Action MP0806,
by the European Research Council under the Atmospheric Research Project No.\
227915, and by a grant from the Government of the Russian Federation under
contract No. 11.G34.31.0048 (NK, IR).
We acknowledge the allocation of computing resources provided by the
Swedish National Allocations Committee at the Center for
Parallel Computers at the Royal Institute of Technology in
Stockholm and the National Supercomputer Centers in Link\"oping.
\end{acknowledgements}

%r e f
\newcommand{\ymonber}[3]{ #1, {Monats.\ Dt.\ Akad.\ Wiss.,} {#2}, #3}

%\vfill\bigskip\noindent\tiny\begin{verbatim}
%$Header: /var/cvs/brandenb/tex/illa/NEMPI_rotation/paper.tex,v 1.143 2012-10-27 05:08:20 brandenb Exp $
%\end{verbatim}

\end{document}